\documentclass[letterpaper]{article} % DO NOT CHANGE THIS
\usepackage{aaai22}  % DO NOT CHANGE THIS
\usepackage{times}  % DO NOT CHANGE THIS
\usepackage{helvet}  % DO NOT CHANGE THIS
\usepackage{courier}  % DO NOT CHANGE THIS
\usepackage[hyphens]{url}  % DO NOT CHANGE THIS
\usepackage{graphicx} % DO NOT CHANGE THIS
\usepackage{natbib}  % DO NOT CHANGE THIS AND DO NOT ADD ANY OPTIONS TO IT
\usepackage{caption} % DO NOT CHANGE THIS AND DO NOT ADD ANY OPTIONS TO IT
\DeclareCaptionStyle{ruled}{labelfont=normalfont,labelsep=colon,strut=off} % DO NOT CHANGE THIS
\usepackage{algorithm}
\usepackage{algorithmic}
\usepackage{caption}
\usepackage{subcaption}
\usepackage{newfloat}
\usepackage{listings}
\floatstyle{ruled}
\newfloat{listing}{tb}{lst}{}
\floatname{listing}{Listing}

\usepackage{longtable}
\usepackage{makecell}
\usepackage{multirow}
\usepackage{tabularx}
\usepackage{booktabs}
\usepackage{tabulary}
\usepackage{comment}
\newcolumntype{C}[1]{>{\PreserveBackslash\centering}p{#1}}
\newcolumntype{R}[1]{>{\PreserveBackslash\raggedleft}p{#1}}
\newcolumntype{L}[1]{>{\PreserveBackslash\raggedright}p{#1}}

\usepackage{xcolor}

\title{Gender differences in online communication: A case study of Soccer}

\author {
     Mariana Macedo\textsuperscript{\rm 1},
     Akrati Saxena\textsuperscript{\rm 2}
}
\affiliations {
    \textsuperscript{\rm 1} Northeastern University London, United Kingdom\\
    \textsuperscript{\rm 2} Leiden Institute of Advanced Computer Science, Leiden University, the Netherlands\\
    mariana.macedo@nulondon.ac.uk, a.saxena@liacs.leidenuniv.nl
}

\begin{document}

%\nocopyright 
\copyrighttext{}%\{\emph{text}\}
\maketitle

\begin{abstract}
Social media and digital platforms allow us to express our opinions freely and easily to a vast number of people. In this study, we examine whether there are gender-based differences in how communication happens via Twitter in regard to soccer. Soccer is one of the most popular sports, and therefore, on social media, it engages a diverse audience regardless of their technical knowledge. We collected Twitter data for three months (March-June) for English and Portuguese that contains 9.5 million Tweets related to soccer, and only 18.38\% tweets were identified as belonging to women, highlighting a possible gender gap already in the number of people who participated actively in this topic. We then conduct a fine-grained text-level and network-level analysis to identify the gender differences that might exist while communicating on Twitter. Our results show that women express their emotions more intensely than men, regardless of the differences in volume. The network generated from Portuguese has lower homophily than English. However, this difference in homophily does not impact how females express their emotions and sentiments, suggesting that these aspects are inherent norms or characteristics of genders. Our study unveils more gaps through qualitative and quantitative analyses, highlighting the importance of examining and reporting gender gaps in online communication to create a more inclusive space where people can openly share their opinions. 
\end{abstract}

\section{Introduction}

Social media platforms, such as Twitter (currently known as X), Instagram, and Reddit, have emerged as significant channels for self-expression. The versatility, convenience, and widespread accessibility of social media encourage people to express and garner reactions from their targeted audience. Fans, sports organizations, and players heavily rely on these platforms for interactive communication and to gather real-time information about ongoing events~\cite{Williams2014,webist16}. It is also likely that breaking news comes first on Twitter than traditional media channels, providing an excellent medium to access instantaneous information from official as well as unofficial sources~\cite{webist16}.

However, on social media, it is often observed that majority groups dominate the communication, and as a consequence, the minority groups do not feel as free to express~\cite{minority_analysis}. Past research has explored gender-related differences in offline as well as online communication spaces, such as Twitter \cite{holmberg2015gender, Garcia_Weber_Garimella_2014}, Stack Overflow \cite{may2019gender}, Wikipedia \cite{man_wikipedia}, and YouTube \cite{youtube_gender}. For instance, Messias et al. \shortcite{messias2017white} showed that, in general, white men are more likely to be followed on social media and acquire higher positions in rankings and ratings on Twitter; only after the top 14\% most followed users, there are higher fractions of women than of men. This can also be translated to specific topics (e.g. Sports, Politics, News, Technology, Industry, and Art)~\cite{nilizadeh2016twitter,manzano2019women} where women tend to be less influential as well as hold lower-ranking/percentage of top influential females. Thus, thoughts and narratives of males linger more on Twitter and are more likely to get traction probably due to this participation imbalance and gender biases and constructs~\cite{Garcia_Weber_Garimella_2014}.

Many social, contextual, and developmental~\cite{Chaplin2015} factors play a role in perpetuating biases and stereotyping in society that is manifested on social media. Some fields and topics such as sports, STEM (Science, Technology, Engineering, and Mathematics) and politics are more likely to be associated with men~\cite{man_wikipedia} facilitating their participation in conversations. Thus, in these contexts, women can feel more hesitant to talk about such topics and take up certain professions due to gender norms and constructs. Skewed narratives and perspectives get reinforced as a consequence and get embedded in subconscious behaviour. To move against this reinforcement, we need to be aware of gender differences to promote more diversity, inclusivity and fairness in online platforms. 
We are interested in analyzing the gender gap in online communication. We focus our analysis on a topic that is widely spoken worldwide: Soccer (or Football). In fact, the last World Cup was watched by more than 4 billion people out of a world population of almost 8 billion \cite{soccernews}, indicating the worldwide importance of the topic. 

We collected soccer-related Tweets for three months for our study using Twitter API. Our dataset includes 7 million English tweets from 2 million unique users and 2.6 million Portuguese tweets from 0.5 million unique users. Our analysis aims to explore gender-based communication differences in both languages. The reason behind choosing Portuguese as a second language is that in Brazil and Portugal, individuals are exposed to soccer from an early age, potentially encouraging people to love the sport and participate more actively than in cultures where soccer is not as popular.

Moreover, some papers found active participation from women in topics related to soccer. \cite{yoon2014gender} showed that the main motivational factor among females to use Twitter is entertainment for leisure time, and not information and fanship. \cite{clavio2013effects} found that female fans were more likely to contribute to sports team feeds on Twitter than men. They rated a variety of informational and commercial functions relating to in-game updates, game results, individual player news, contests, giveaways, and ticket discounts, more than men. This study also found that female fans were significantly more active on social media using smartphones while being at the stadium and more likely to respond to Tweets sent out by team Twitter feeds than male fans. In this paper, we investigate these ideas further to understand how women as a minority might have a space to be active in a male-dominant topic such as soccer and how this impacts communication behaviour.

Women make up almost half of the population and 38\% of the soccer fan base \cite{lange_2020}, and yet we only observed 18\% of tweets were from women. Across our dataset, women were found to be much less likely to talk about soccer compared to men and exhibited more retweeting behaviour. This can be an evidence of a starker gender gap. Through our study, we investigate various ways in which men and women behave differently in terms of communicating, such as type of content, frequency of content, and behaviour in a community. We conduct three levels of analysis: text-level, network-level, and event-level, to gain insights and answer the research questions. We observe that the Portuguese network has lower homophily than the English network, showing that more interactions between men and women might be related to the popularity of a topic, making it more inclusive. However, in both languages, we observe that women show intense emotions as men and that the language of the tweets plays a small role in gender differences related to emotions and sentiments.

Here, we wish to drive communication about this topic to encourage a healthier and equitable online environment. We look into the factors that make a space comfortable for everyone to express freely without fearing judgment. UNESCO launched a \#ChangeTheGame campaign to encourage gender equality in sports as they believe that football can act as an effective medium to reduce the gender gap and empower women worldwide~\cite{unesco_2020}. Analyzing the evolution of football-related communication can help us to create a safe space online through interventions and encourage participation from everyone. This might ignite a change from the grassroots level and amplify women's involvement in the game across all levels, right from players, fans, sport experts, coaches, and positions in governance organizations. While recognizing privacy, ethical and computational challenges, our current analysis is limited to binary genders, and we acknowledge the importance of expanding to a broader gender spectrum in future studies.

Our paper is divided into four parts. In the next section, we present our dataset, followed by our findings. The paper is concluded with a discussion including future directions.

\section*{Soccer-Tweet Dataset}

\begin{figure}[]
    \centering
    \includegraphics[width=\linewidth]{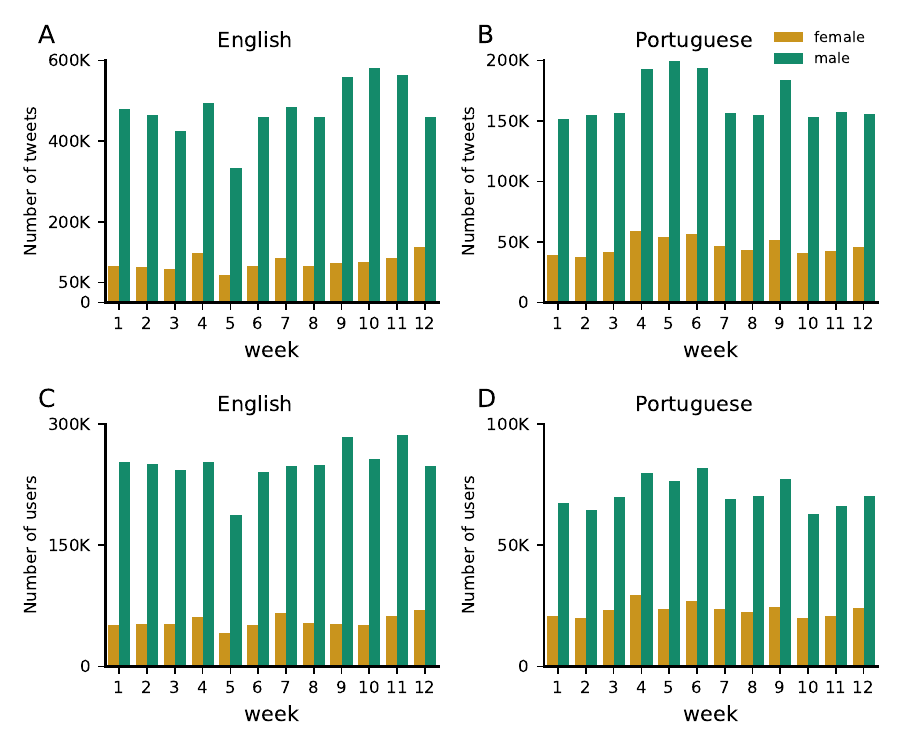}
    \caption{Number of tweets and users per week disaggregated by gender}
    \label{fig_Datastats}
\end{figure}

We collected data from Twitter over a span of three months (March 7 to June 5, 2022) using the following pattern: ``\#Soccer OR football OR futebol OR \#soccer OR \#football OR \#futebol OR \#futball OR \#futbol OR \#fifa OR \#CBF OR \#brasileirao OR \#CopaIntelbrasDoBrasil OR \#Libertadores OR \#Sudamericana OR \#CampeonatoDoBrasileiro OR \#CopaAmerica OR \#ConfedCup OR \#UEFACup OR \#Supercup OR \#WorldCup OR \#LaLiga OR \#CopadelRey OR \#CoppaItalia OR \#CoupedeFrance OR \#premierleague OR \#championsleague OR \#FACUP OR Confederations Cup OR \#confederationscup OR \#ConfedCup OR Copa del Rey OR EFL cup''. We carry out the gender detection using Genderize~(\citeauthor{genderize}) and Namepedia~(\citeauthor{namepedia}). After the gender detection, we discard the tweets whose users could not be assigned a gender (male or female). After these steps, in English (also refereed by \textbf{\textit{en}}), we were left with 6,957,598 tweets: 5,767,122 tweets by males and 1,190,476 tweets by females. Tweets by females had a composition of 17.11\% as compared to 82.89\% by males. In the time frame of 3 months, we had 421,996 unique female users and 1,556,818 unique male users. Similarly, in Portuguese (also refereed by \textbf{\textit{pt}}), we had 2,572,247 tweets, where 2,011,286 tweets were from males, and 560,961 (21.81\%) tweets were from females. We had 365,045 unique male users and 148,539 (28.92\%) unique female users. For both languages, men have a majority in terms of users and posting tweets overall and per week (shown in Figure~\ref{fig_Datastats}).

\subsubsection*{Retweeting Behavior}

We also separated retweets and replies to examine whether they have a distinct pattern from the original posted tweets. We observed in English that 68.91\% of the female tweets were retweets compared to 60.25\% for males, and in Portuguese, 59.07\% of the female tweets were retweets compared to 55.08\% for males. For the replies, we observe in English, 15.87\%(23.21\%) of female(male) tweets, and in Portuguese, 20.67\%(37.87\%) of female(male) tweets. Thus, women tend to retweet more than men, and the opposite is observed for replies.

\section{Our Findings}

We divide our analysis into two parts - analyzing the content of the tweets and analyzing the networks generated from the interaction between users through retweets and replies. 

\subsection{Text Analysis}

We first analyze the tweet text using emojis, hashtags, sentiments, emotions, and other linguistic attributes.

\subsubsection*{Emoji Usage Analysis}
A picture is worth a thousand words - The emojis we use today in online communication are more than a tiny picture; they symbolize emotions. Emoji usage has increased rampantly so much that the Oxford Dictionary declared the emoji face with tears of joy (emojifacewithtearsofjoy{}{}) as the word of the year for 2015 \cite{2015_emoji_of_the_year_oxford}. The scientific research community has also shown interest in analyzing emoji usage patterns in communication~\cite{https://doi.org/10.48550/arxiv.2105.05887, https://doi.org/10.48550/arxiv.2105.03160, https://doi.org/10.48550/arxiv.2105.00846, Chen_2018}. 

In this work, we use the Emot \cite{Emot} library to extract emojis from the tweet text. After extraction, we analyze how emoji usage varies by the two genders. In the English female tweet data, around 32\% tweets contained at least one emoji as compared to 29.20\% of tweets by males. However, if we remove the retweets, quoted tweets, and replies, and only look at the original tweets, we find that 29.85\% of tweets by females had emojis, while only 21.42\% of original male tweets had emojis. The results indicate that emoji usage tends to be more prevalent among females in both languages (Table \ref{emojitable}). This finding also aligns with the study by Chen et al.\cite{Chen_2018}, where they report that female users include emojis in a larger proportion of text messages as compared to men. Figure \label{fig_top_20_emojis} shows top-20 emojis used by both genders and the percentage of tweets that used that emoji. This gives us insight into the type of emojis that are being used by males and females in the context of soccer communication.

\begin{figure}[t]
\centering
\includegraphics[width=\linewidth]{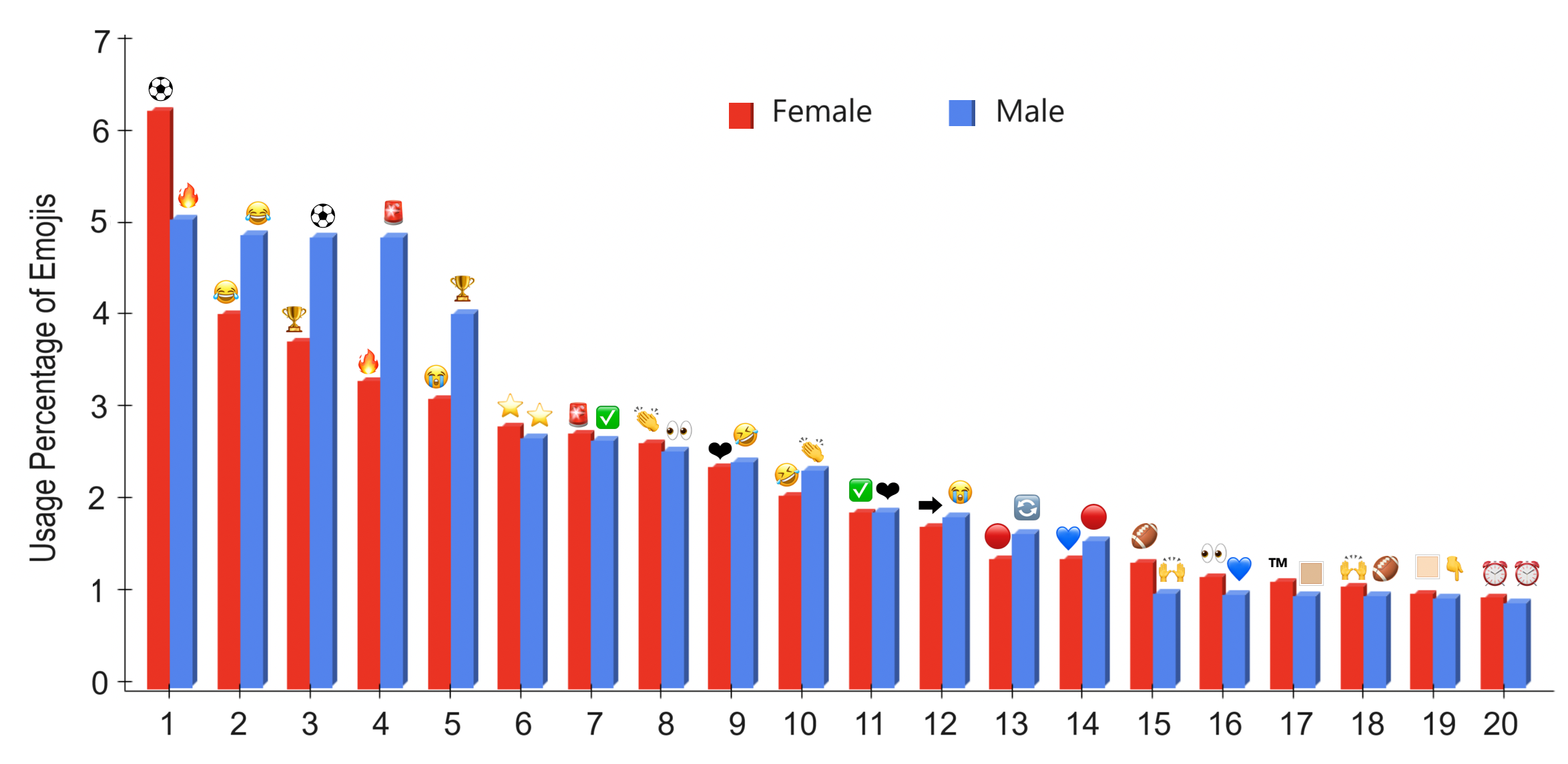}
\caption{Top 20 Emojis and their usage proportion by males and females}
\label{fig_top_20_emojis}
\end{figure}

\begin{table}[h]
    \centering
    \caption{Emoji Usage by females and males} %Average (standard variation) values of Network Portrait divergence (NPD).
    \resizebox{0.45\textwidth}{!}{  
        \begin{tabular}{lllcc} \toprule
             & gender & tweets & \thead{\% with emoji} & \thead{ average \# of emojis}\\ 
             &  &  & & \thead{(including 0s)}\\ 
            \midrule
            
            \multirow{4}{2em}{En} & female & original & 29.85\% & 2.42 (0.72) \\
            & & all & 31.91\% & 2.35 (0.75) \\
            & male & original & 21.43\% & 2.44 (0.52) \\
            & & all & 26.28\% & 2.32 (0.39) \\
            \midrule

            \multirow{4}{2em}{Pt} & female & original & 22.59\% & 3.11 (0.70) \\
            & & all & 22.43\% & 3.56 (0.80) \\
            & male & original & 17.79\% & 2.75 (0.48) \\
            & & all & 16.19\% & 2.78 (0.45) \\
          
            \bottomrule
        \end{tabular}
    }
    \label{emojitable}
\end{table}

Emojis are used further in sentiment and emotion analysis. We keep the emojis in their raw format while computing the sentiment using VADER API~\cite{Hutto_Gilbert_2014} as the model is capable of handling emojis. For emotion analysis, the emojis were converted to their equivalent emotion text so that they could be captured by the NRC-Emotion Intensity Lexicon API \cite{LREC18-AIL}.

\subsubsection*{Hashtag Usage}

Hashtags are a pivotal part of communication on Twitter and help people follow the content related to a topic easily using its specific hashtags (\citeauthor{twitter_hashtag}). In our dataset, from 7.53 million tweets, there are almost 3 million hashtags, and 20.77\% of the tweets had at least one hashtag. 23.12\% of the tweets with at least one hashtag were from females, as compared to 20.28\% of tweets from males. This shows marginally higher hashtag usage behaviour by women as compared to men. However, both women and men are more likely to use one hashtag per tweet than several ones. The average number of hashtags across genders is similar; for females(males), it is 0.8587(0.8691) in English and 0.1686(0.1518) in Portuguese. 

\subsubsection*{Sentiment Analysis}

\begin{table}[t]
    \centering
    \caption{Fraction of tweet for each sentiment category} %Average (standard variation) values of Network Portrait divergence (NPD).
    \resizebox{0.3\textwidth}{!}{  
        \begin{tabular}{llccc} \toprule
            &  &  \multicolumn{3}{c}{Sentiment} \\ 
             & Gender & Pos & Neg & Neu\\ 
            \midrule

            \multirow{2}{2em}{En} & female & 0.55 & 0.21 & 0.24 \\
            & male & 0.55 & 0.21 & 0.24 \\
            \midrule

            \multirow{2}{2em}{Pt} & female & 0.10 & 0.05 & 0.85 \\
            & male & 0.09 & 0.05 & 0.86 \\
          
            \bottomrule
        \end{tabular}
    }
    \label{tab:sentimenttable}
\end{table}

\begin{figure}[]
\centering
\includegraphics[width=0.9\linewidth]{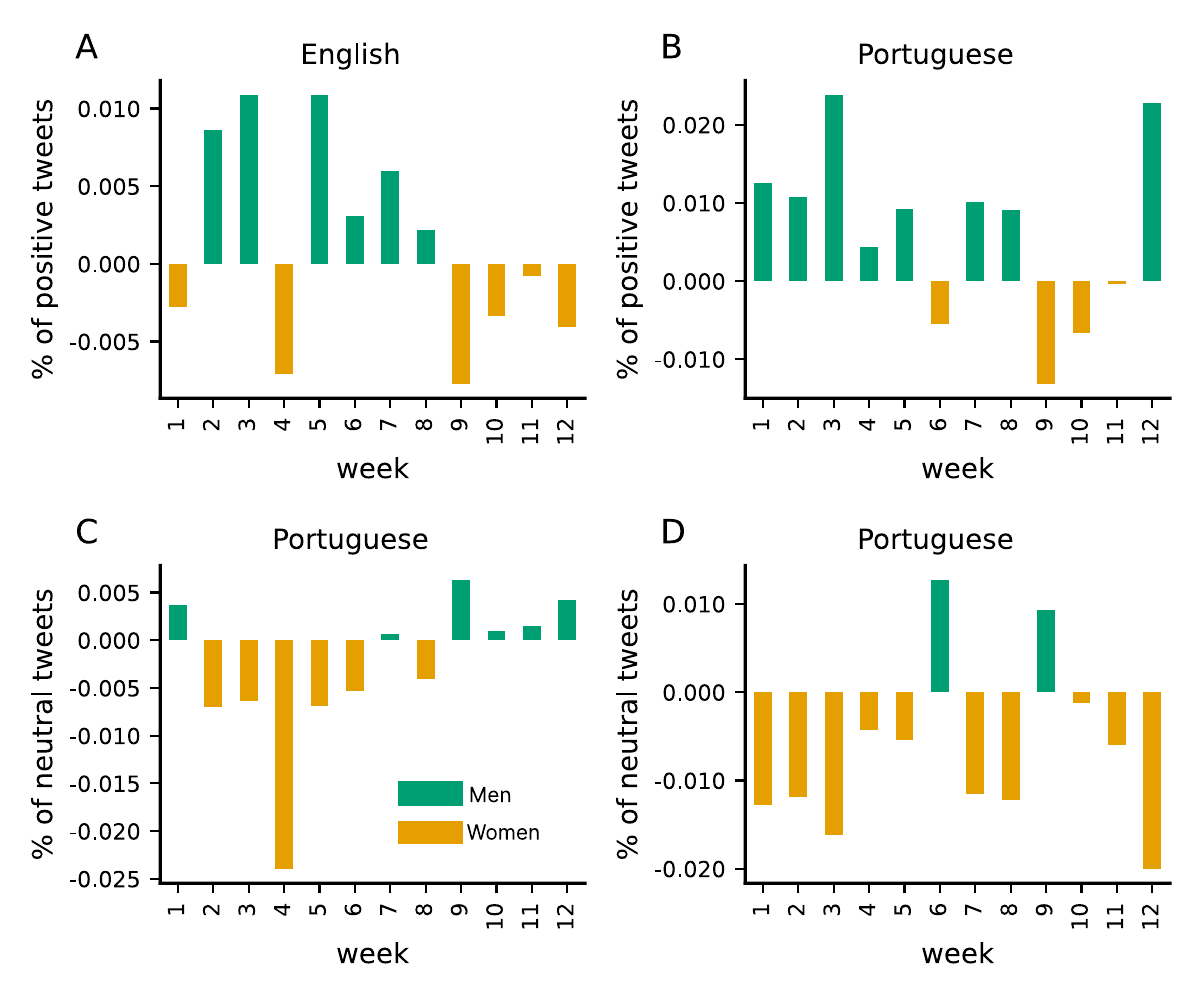}
\caption{Men have a higher percentage of positive tweets, and women post more neutral content.} %women tend to be more neutral
\label{fig:sentiment_gender}
\end{figure}

Analyzing sentiments \cite{saxena2022recent} from content helps us to understand public opinions and preferences at a large scale for different topics, such as events, policies, and laws~\cite{article_italian_tweets, cheng2021evaluation, article_twitter_opinion_inversion, saxena2022introduction}. We use VADER (Valence Aware Dictionary for Sentiment Reasoning) \cite{Hutto_Gilbert_2014} to study how the tweets by men and women differ sentimentally. VADER is widely used for social network data due to its ability to handle texts with emojis, emoticons, extra punctuations, negations, use of contractions, randomized capitalization, use of degree modifiers, sentiment-laden slang words, and acronyms. VADER is a lexicon and rule-based tool to get sentiment scores across categories `Positive' (compound score $\geq$ 0.05), `Negative' (compound score $\leq$ -0.05) and `Neutral' (compound score between -0.05 and 0.05)~\cite{Hutto_Gilbert_2014}. Therefore, each tweet was classified into one of the three categories: positive, negative and neutral.

Overall, the average sentiment scores of both genders are very similar (Table \ref{tab:sentimenttable}), but when we disaggregate the tweets per week, we observe that men post more positive tweets and women are more neutral (Figure \ref{fig:sentiment_gender}); for negative tweets, we did not observe any dominance.
We also further study how different genders show sentiments for events, and the results are interesting, as shown in Figure \ref{appendixfigsentiment_distribution} in Appendix A. We observe that whenever an event happens, females show more intense sentiments than males. We identified some events that were on days with intense sentiments for English as an example. The events corresponding to intense positive emotion include - \textit{Real Madrid won the La Liga for the 35th time and Champion League final won by Real Madrid}, and for negative sentiments, they include - \textit{Algeria Cameroon Match Controversy, Dwane Haskins, Americal Pittsburgh Steelers quarterback died in an accident and Ukraine lost making way for Wales in the world cup}. Similar results were observed for Portuguese. 

\begin{figure}[t]
\centering
\includegraphics[width=\linewidth]{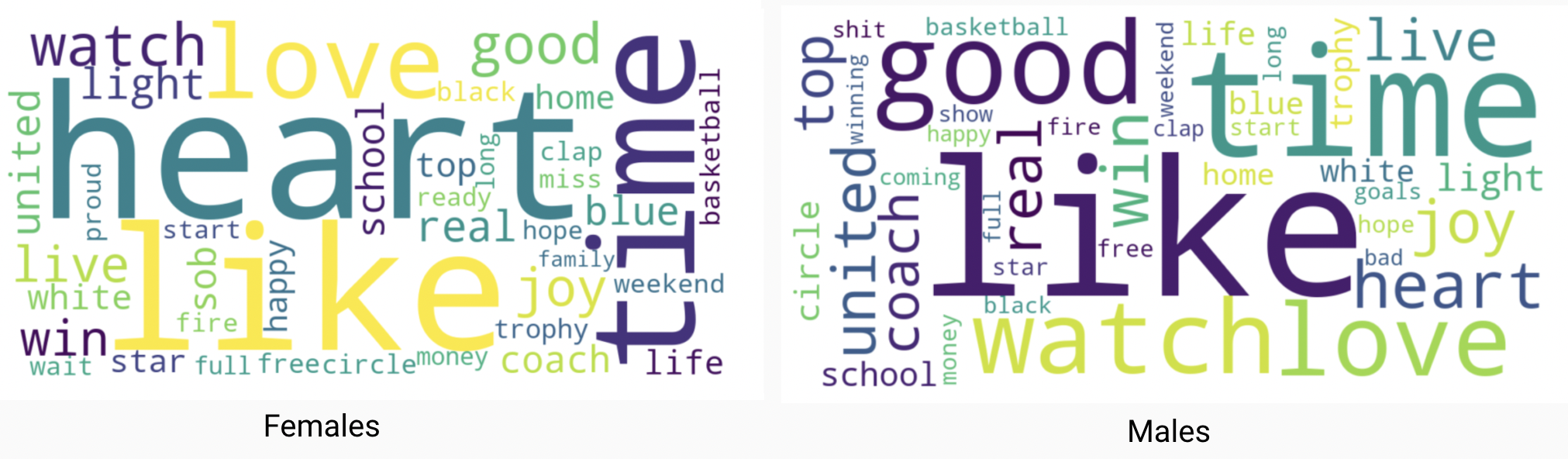}
\caption{Word cloud of words used to show emotions (for English)}
\label{fig_word_cloud_of_emotions}
\end{figure}

\begin{figure}[h]
\centering
\includegraphics[width=\linewidth]{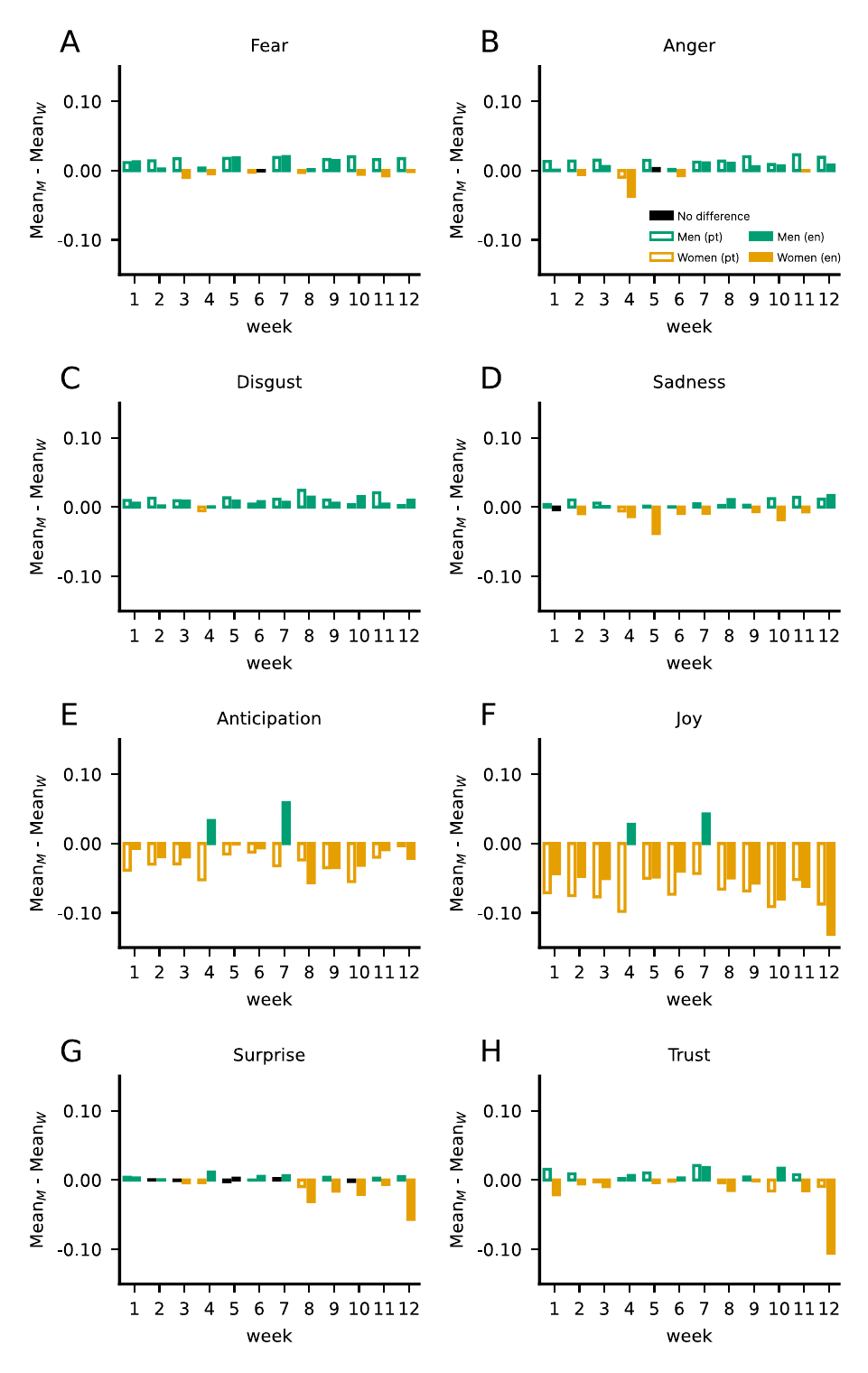}
\caption{Gender differences from the emotion usage from tweets over time }
\label{fig:emotions_diff}
\end{figure}

\subsubsection*{Emotion Analysis}

Figure~\ref{fig_word_cloud_of_emotions} shows the word clouds of the most frequently used words to convey emotions. Females tend to use the words ``Heart'' and ``Love'' more frequently than males. Men tend to use the words ``Like" and ``Good'', which display less intense emotions. These choices of words by males indicate a proclivity towards conveying emotions that are less intense and perhaps more casual in nature. 

In general, the most predominant emotions for both males and females for the topic of soccer were Joy ($\approx32\%$ tweets), Anticipation ($\approx23\%$ tweets), and Trust ($\approx22\%$ tweets). To understand the gender differences, we compared the distribution of emotions per week using a bootstrapping with 80\% subsamples simulated for 1000 iterations (Figure \ref{fig:emotions_diff}). These results show that the intensity of joy and anticipation is higher for females as compared to males, and the opposite happens to the emotions of fear, anger, and disgust. For the emotions of surprise and sadness, there is no clear predominance. 

Overall, the analyses of the word clouds and emotions suggest that women and men tend to express themselves differently. They tend to use different words and expressions, and the intensity with which these words are expressed tends to be different, ultimately shaping how emotions are conveyed in communication. 

\subsubsection*{Content Analysis} 
\begin{figure}[t]
\centering
\includegraphics[width=\linewidth]{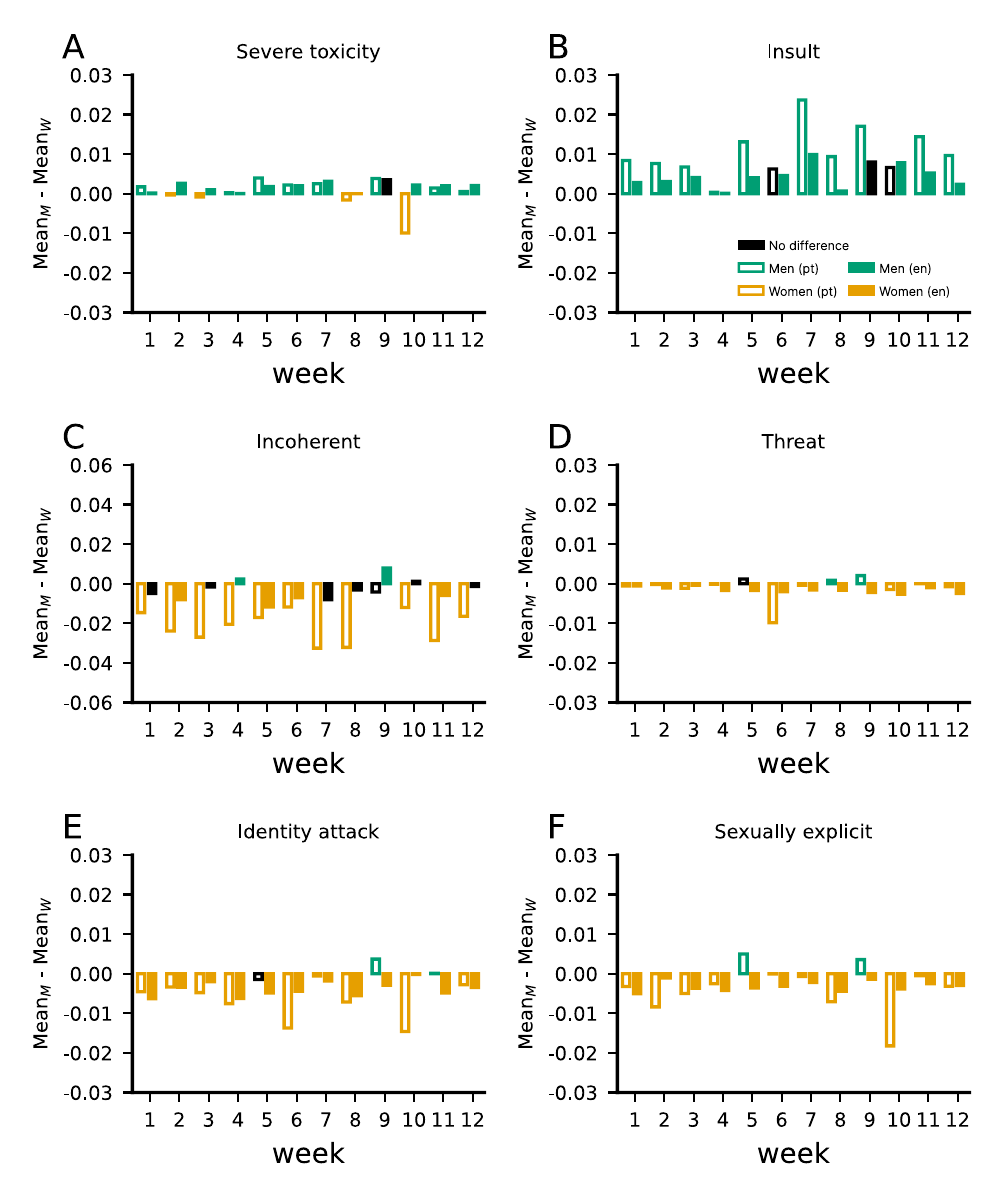}
\caption{Statistical difference in writing style based on toxicity, insult, incoherent, threat, identity attack, and sexually explicitness} 
\label{Figure_toxicity_2}
\end{figure}

From the analyses of sentiments, we noticed that irony and injury were not well detected and that the results from the emotion of joy and the sentiment of positivity were not aligned. Therefore, we decided to analyze the content in relation to possible toxicity and abusive behaviour online. We used Google's Perspective API \cite{Hutto_Gilbert_2014} to estimate the levels of severe toxicity, insult, attack on the author, threat, identity attack, and sexually explicit behaviour. We used the same bootstrapping technique of 80\% subsampling and 1000 repetitions, and plotted the difference between the average values of each metric for each gender (Figure \ref{Figure_toxicity_2}). The values tend to be small (around 0.13), and we just analyze the gender differences. We observe that men tend to express more severe toxicity, insult and attack on author, and women tend to express more threat, identity attack and sexually explicit behaviour. We can then conclude that there is still a need to reduce abusive communication. Results for several other metrics are discussed in Appendix B. 

\subsection{Network-Analysis}

We build weighted cumulative networks (from the starting week till the last week) from the communication between the users (nodes), where the link weight is the number of times two users interacted with each other by replying or retweeting information. There are three types of networks for each language based on (i) retweets, (ii) replies, and (iii) both retweets and replies (mentioned as the combined network based on all communications). The size of the networks are - in English, there are 148,184 females (18.94\%) and 634,366 males, and in Portuguese, there are 50,354 females (25.85\%) and 144,441 males. In total, we have 4,308,009 interactions in English, where 1,348,557 are replies and 2,959,452 are retweets, and in Portuguese, we have 2,117,105 interactions, where 986,288 are replies and 1,130,817 are retweets. We will analyze these networks to understand how information is shared for each language based on their structure and flow.

\subsubsection*{Communication Patterns}
We first observed that the network from the communication across languages tends to be different ($NPD>0.90$, results in Table~\ref{tab:portrait}) by computing the Network Portrait Divergence (NPD) between networks that estimates how different two networks are in relation to their structure (0 to 1 corresponds to being similar to different)~\cite{bagrow2019information}. The NPD method compares networks using the network portrait that is a $(l, k)$ size array containing the number of nodes who have $k$ nodes at a distance $l$. Therefore, it provides an information-theoretic interpretation for compared networks based on the structures of all scales and is well-generalized for weighted networks. We observe that the networks tend to be different in structure, except when we compare the two networks built solely from the retweets, then NPD tends to be close to 0.4. This means that the replies play a major role in making these interactions different. Next, we use NPD to compare the structures from both networks, but now we compare how women and men interact. Overall, women and men tend to emerge in different networks (values of NPD closer to 1). However, we see that for replies in English, the network structure tends to be similar for women and men, indicating that the networks built from Portuguese tweets are more different in structure than the ones from English. 

\begin{table}[h]
    \centering
    \caption{Network Portrait divergence (NPD) average (standard variation) values from a bootstrap technique. We built networks from 80\% of data (1000 times).} 
    \resizebox{0.35\textwidth}{!}{  
        \begin{tabular}{lll} \toprule
            \thead{Tweets} & \thead{Comparison} & \thead{NPD} \\ 
            \midrule
            
            \multirow{3}{2em}{all} 
            & across languages & 0.90 (0.007) \\ 
            & English, across gender & 0.97 (0.01)  \\ 
            & Portuguese, across gender & 0.99 (0.001)  \\ 
            \midrule
            
            \multirow{3}{2em}{retweets} 
            & across languages & 0.46 (0.04) \\ 
            & English, across gender & 0.39 (0.01)  \\ 
            & Portuguese, across gender & 0.41 (0.02)  \\ 
            \midrule
            
            \multirow{3}{2em}{replies} 
            & across languages & 0.91 (0.007) \\ 
            & English, across gender & 0.28 (0.02)  \\ 
            & Portuguese, across gender & 0.94 (0.001)  \\ 
            \bottomrule
        \end{tabular}
    }
    \label{tab:portrait}
\end{table}

\begin{figure*}[t]
\centering
\includegraphics[width=0.96\linewidth]{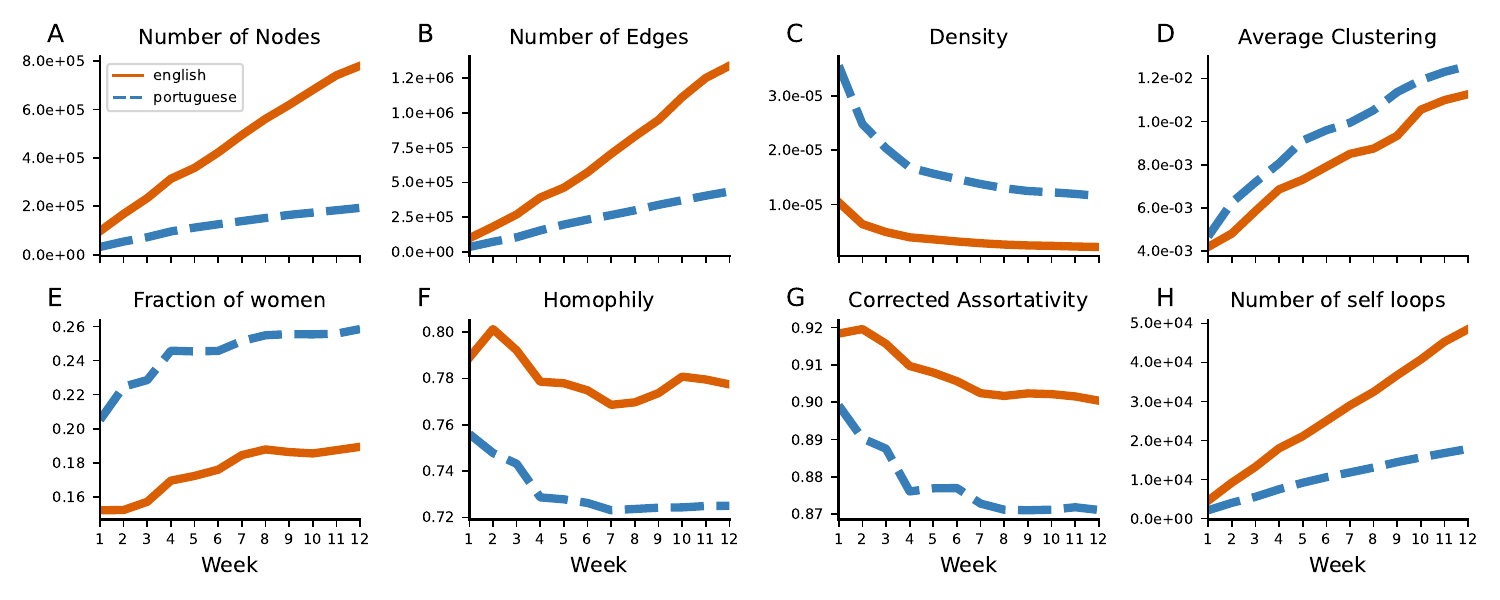}
\caption{Network metrics computed from the combined network of retweets and replies for Portuguese and English.} 
\label{fig:network_overall_characteristics}
\end{figure*}

\begin{figure*}[h]
\centering
\includegraphics[width=0.9\linewidth]{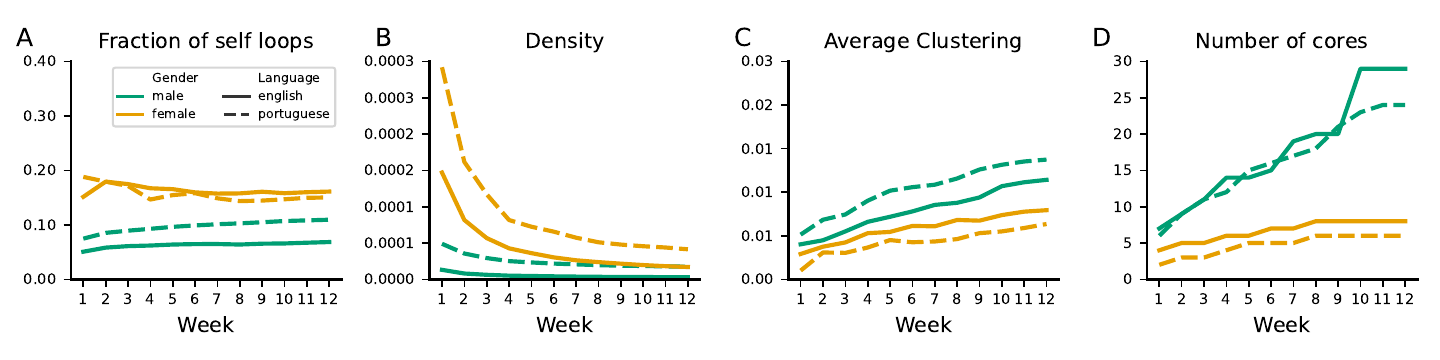}
\caption{Network metrics for the combined network (retweets + replies) disaggregated by gender for Portuguese and English.}
\label{fig:network_gender}
\end{figure*}

\begin{figure*}
    \centering
    \includegraphics[width=0.95\linewidth]{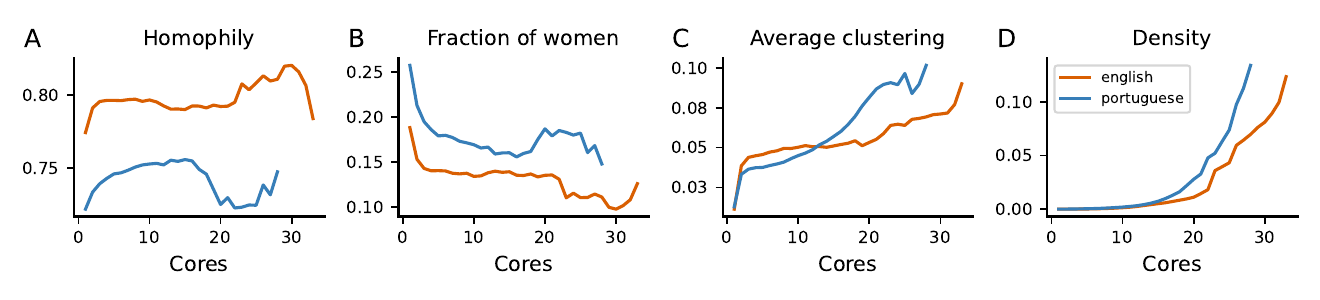}
    \caption{Evolution of the metrics as we go from the core 1 to the maximum number of cores.} %unweighted network
    \label{cores:evolution}
\end{figure*}

\begin{figure*}
    \centering
    \includegraphics[width=0.95\linewidth]{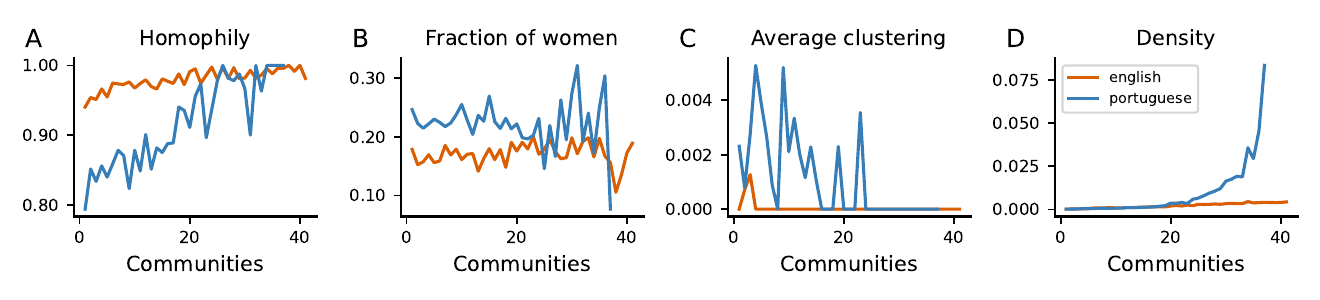}
    \caption{Evolution of the metrics as we go from the biggest (labeled as 1) to the smallest community. } %unweighted network
    \label{communities_evolution}
\end{figure*}

\subsubsection*{Network Characteristics} 
Let us investigate further the characteristics of these networks. We compare the average values of multiple metrics computed from the networks built from all the tweets in English and Portuguese (Figure \ref{fig:network_overall_characteristics}). 
From the cumulative evolution over 12 weeks, we observe that the number of nodes and edges is higher for English, and the density and average clustering tend to be smaller. We argue that people tweeting in Portuguese might belong to more concentrated regions than in English (worldwide spoken language). Therefore, soccer might come as a more popular and well-connected local phenomenon in Brazil and Portugal, but for English tweets, we look at several communities together, indicated by the higher connection between small groups (closer average clustering values) but not to the overall network (much smaller density).

Even though English tweets tend to come from people from different regions, the fraction of women does not scale at the same rate as Portuguese. A much smaller community coming from Portuguese tweets shows a much larger fraction of women. In general, women tend to be a minority in sports-related topics~\cite{farrell2019exploring,thelwall2019she}, but the popularity of soccer in smaller communities (Portuguese ones) might encourage women to participate more in the topic. This higher participation translates into more active participation across genders. Women from Portuguese tweets tend to interact more with men than from English (lower values of gender homophily for Portuguese). To make sure that these differences are not a group size effect, we compute the assortativity of both networks, and we still observe the same patterns. The assortativity is computed using the adjusted nominal assortativity, i.e., defined explicitly for networks having asymmetric mixing of nodes from different groups (genders in our case)~\cite{karimi2023inadequacy}. Thus, the increase in the fraction of women in the Portuguese network over weeks impacts the homophily values, and as there are more women, the homophily and assortativity are further reduced. 

Next, we look at the self-loops in the networks (Figure \ref{fig:network_overall_characteristics}.H). We see that English tend to have more self-loops than Portuguese, as is expected by their group size. However, when we disaggregate these self-loops by gender (Figure \ref{fig:network_gender}.A), we see that around 10\% of the edges coming from men's interactions are self-loops which is half of what we see from women's interactions. This indicates that women in our sample tend to retweet and reply more to their own content than men. By disaggregating other network metrics based on gender, we also see that women have denser networks, but men tend to have higher average clustering and higher core numbers. This indicates that women tend to be more connected in general, but men tend to act more as hubs and acquire more central positions. 

The core number of a node shows its influential power, and the nodes at the very internal core (corresponding to the highest core number) are the top influential nodes \cite{kitsak2010identification}. The highest number of cores for English is 33, and for Portuguese is 28. At these cores, we analyze the networks in Portuguese (169 people) and English (214 people), and we found a women ratio of 14.79\% and 12.61\%, respectively. Even though women have a higher ratio of most active users in Portuguese than in English, the difference between the women ratio of the overall network ($f_{w,pt}=25.85\%$, $f_{w,en}=18.94\%$) and the core is decreased by 11.06\% and 6.33\%, respectively, indicating that women are not as influential as men in both languages.

\subsubsection{Meso-scale Structures}

We further study the organization of nodes in the network based on meso-scale structures, including core-periphery and communities \cite{saxena2016evolving}. 
In Figure ~\ref{cores:evolution}, we plot the evolution of metrics over cores in both networks based on all kinds of tweets; English has a higher core number than Portuguese. We identify the core-periphery using the k-shell decomposition method \cite{kitsak2010identification}. We first study how homophily varies as we move from periphery to core to understand the connection pattern of different genders at different influential levels. 
Many previous works have shown that women do not acquire top positions in the network due to the glass-ceiling effect \cite{avin2015homophily, stoica2018algorithmic}. Figure~\ref{cores:evolution} shows the fraction of females in different layers as we move from periphery to core in both the networks, and we observe that when a core of Portuguese reaches a similar fraction of women inside it compared to English, they have a much lower homophily than in English. We observe that homophily increases as we move from periphery to core, equal to 15 for both languages, a fraction of women decreases, and density continues to be similar across languages. However, the average clustering for Portuguese increases much more than for English, probably due to the fact that the people in the network belong to a specific region. 

Next, we study the embedding of females and males in different communities; the results are shown in Figure~\ref{communities_evolution}. Communities are extracted using the Leiden Community detection method \cite{traag2019louvain}. Analyzing the network of English tweets reveals that homophily increases as the group size decreases, and the biggest community reflects the smallest homophily value. This also happens in the Portuguese network, but the variance is much higher. English network is more homophilic (as shown in Figure~\ref{fig:network_overall_characteristics}), and this trend extends to the identified communities. The increase of homophily does not correlate with the values of the fraction of women and average clustering, but for Portuguese, the density increases as the group size decreases and the homophily increases.

\section{Discussion}

In this paper, we study Twitter communication about soccer (or football) in English and Portuguese from a gender perspective, as it is highly popular and the most played sport in the world. While existing works \cite{nilizadeh2016twitter,manzano2019women,amarasekara2019exploring} emphasize gender gaps in communication within specialized fields like STEM and politics, discussing such topics requires specific training and understanding. In contrast, our focus is on topics that encourage people to interact regardless of their expertise, making them more accessible. Examining the extent to which women and men communicate differently about Soccer on Twitter is not only intriguing but also holds significant importance in understanding gender-centric communication dynamics. Our findings aim to expand the understanding of how the popularity and familiarity of a topic influence communication, and what are the gaps where interventions can be tackled. 

We collected soccer-related Twitter data for three months (March 7 to June 6, 2022) and identified the genders of users using Genderize and Namepedia. The scope of our study is confined to binary gender, and tweets without the specified gender of their authors were excluded from further analysis. The final dataset contains 7 million tweets in English (from 2 million users) and 2.5 million tweets in Portuguese (from 0.5 million users). We then analyzed the content of the tweets and interaction patterns from a gender perspective to observe how a male-dominated environment impacts communication. We found that women tend to be a minority in participation and representation, but they express more sentiments and emotions. 

There are several similarities between women and men, such as the similar use of the top 20 most frequently used emojis and the usage of hashtags in the tweets. However, women use more emojis than men, though this does not translate to the usage of hashtags. Men post more positive tweets, and women post more neutral tweets. Gender differences in negative tweets did not show a consistent result. This might be due to the diversity of negative emotions that can be represented by the class ``negative''. We see that when analysing abusive behaviour, women and men can display ``negative'' tweets; that is hard to be captured in an aggregated manner.

Then, we analysed further the emotions detected from the tweets. Women tend to express higher levels of joy and anticipation than men in both languages, and disgust, anger, and fear tend to be more gender-neutral, with slightly higher levels for males. Women display intense sentiments and emotions for any event as compared to men. Besides these, tweets posted by men have higher toxicity, profanity, insulting text, abusive language, and attacking writing style. However, women's tweets are more incoherent, sexually explicit, and have higher identity attacks. Interestingly, we did not find any notable emotional difference based on gender between English and Portuguese, and emotional responses appear to be unaffected by the overall network structure.

We further constructed retweet, reply, and the combined networks based on Portuguese and English tweets. The reply and combined networks are statistically different (computed using NPD) across genders, highlighting a significant difference in communication across genders based on replies. However, the retweet networks across languages and genders are not very different (NPD score is around 0.4). The Portuguese network exhibits a higher proportion of women in interactions and lower homophily compared to the English network. This difference could be attributed to a regional focus in Portuguese tweets, mainly from Brazil and Portugal, where soccer is highly popular. Women's networks in Portuguese show denser structures, with higher average clustering and lower assortativity than those of men. 

Women are less influential, and the ratio of women reduces from the periphery to the core of the network; the fraction is decreased by 11.06\% and 6.33\% in Portuguese and English, respectively. Small communities in both languages are more homophilic. In Portuguese, bigger communities are less homophilic, and homophily reduces with the community size, given that the fraction of women is maintained in most of the communities. However, the difference in homophily across communities is not very significant in English. 

Our observations indicate a strong correlation between a topic's popularity and lower homophily in the network, suggesting a safer space for communication between different genders. This, however, does not influence how individuals express their emotions. The significant communication gap highlighted in this study is eye-opening, and it emphasizes the need to focus on bridging it and creating online safe spaces for discussions on such topics. While we do not propose intervention methods in this work, we believe that the observed outcomes can serve as a foundation. In the future, we would like to study further the correlation of popularity and ease of topics with the homophily of their network and writing style. If similar results are observed, then one can use awareness and education programs to bridge these gaps. Similar techniques could be used for topics that are very domain-specific, such as STEM.

We acknowledge some limitations of this work. First, Twitter's user base may not represent a comprehensive picture of the world as it is predominantly used by specific groups, such as white, male, and middle-upper class people, which can impact the generalizability of results. Nevertheless, studying the use of online social media, such as Twitter, can help us to make these platforms more inclusive. Secondly, some methodologies, such as the API used for sentiment analysis, still needs improvement and refinement. In our case, many tweets that were related to offences and ironies were classified as positive. To address this, we conducted analyses from various perspectives that complementarily can collaborate to enhance results' robustness. For instances, results related to emotions were manually checked, and tend to be more accurate

Our study goes beyond to explore gender variations in communication patterns, highlighting potential misperceptions of free speech on social media. Despite soccer being more popular among Portuguese speakers, we observe similarities (such as influential position for women) as will as variations (such as difference in network structures) in communication patterns across languages. In the future, we plan to further investigate the explainability of patterns extracted from the networks and their communities. Additionally, we plan to compare these patterns with a with a sport or activity predominantly followed by women, such as ballet. We posit that different sports may influence people to communicate in distinct manners. 

\bibliography{aaai22.bib}

\section*{Appendix A}

\begin{figure*}[th!]
\centering
     \begin{subfigure}[b]{\textwidth}
         \centering
         \includegraphics[width=\textwidth]{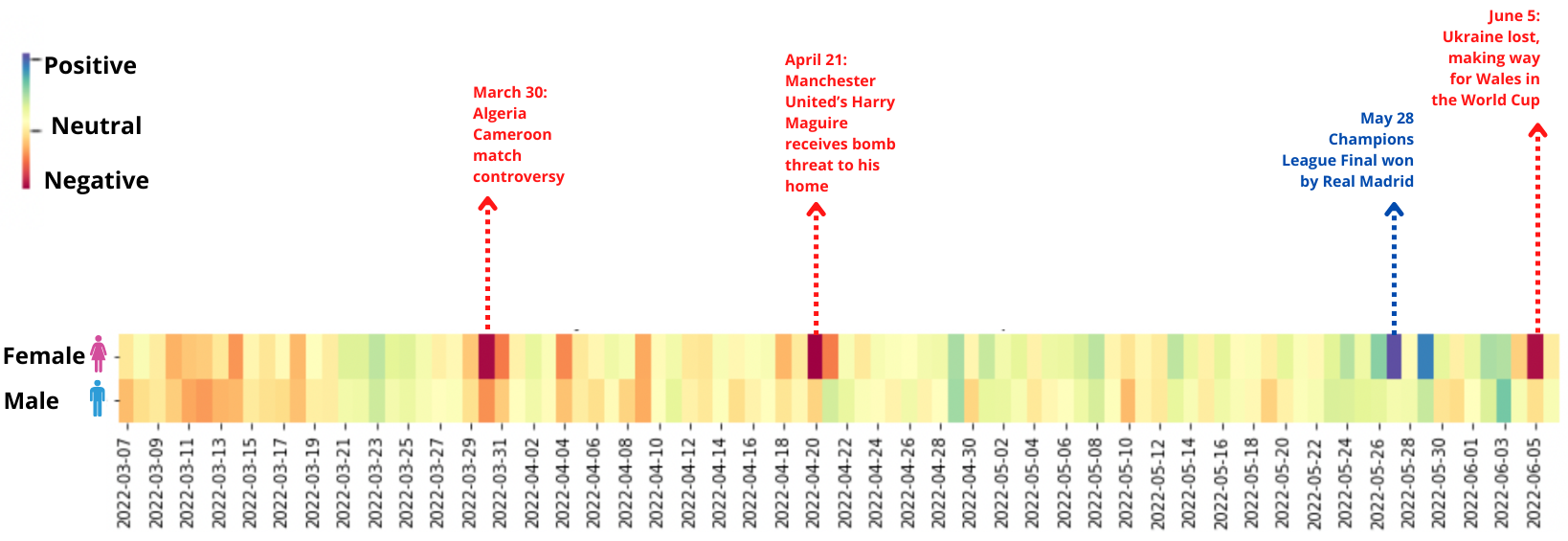}
         \caption{sentiment compound score}
         \label{fig:y equals x}
     \end{subfigure}
     \hfill
     \begin{subfigure}[b]{\textwidth}
         \centering
         \includegraphics[width=\textwidth]{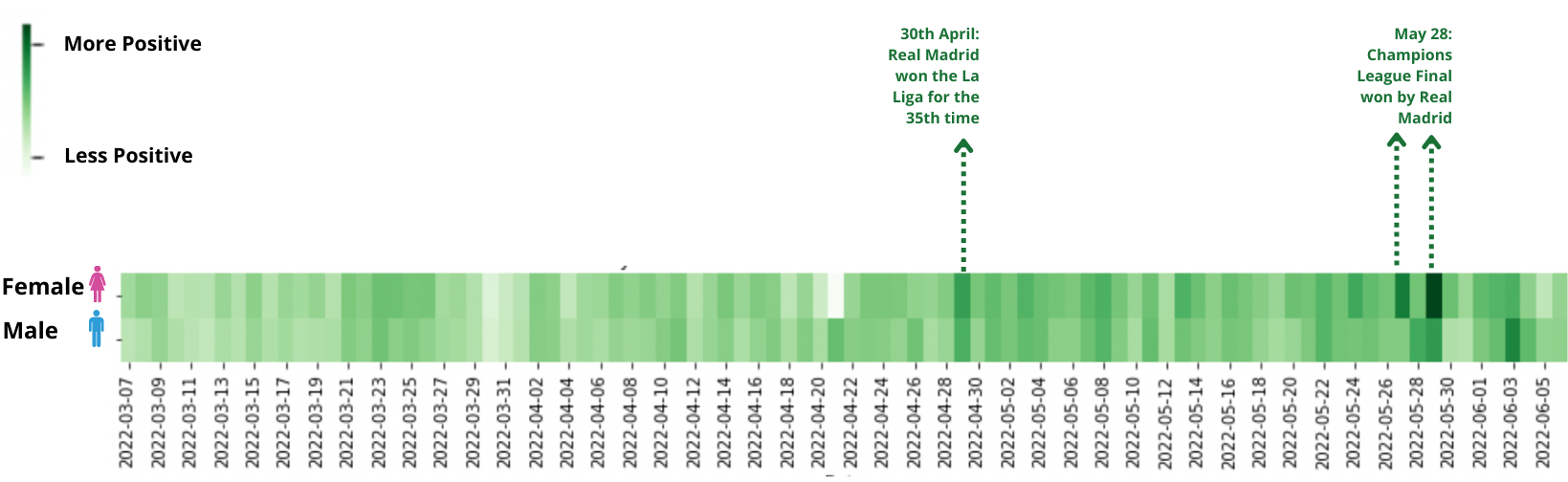}
         \caption{positive sentiment}
         \label{fig:three sin x}
     \end{subfigure}
     \hfill
     \begin{subfigure}[b]{\textwidth}
         \centering
         \includegraphics[width=\textwidth]{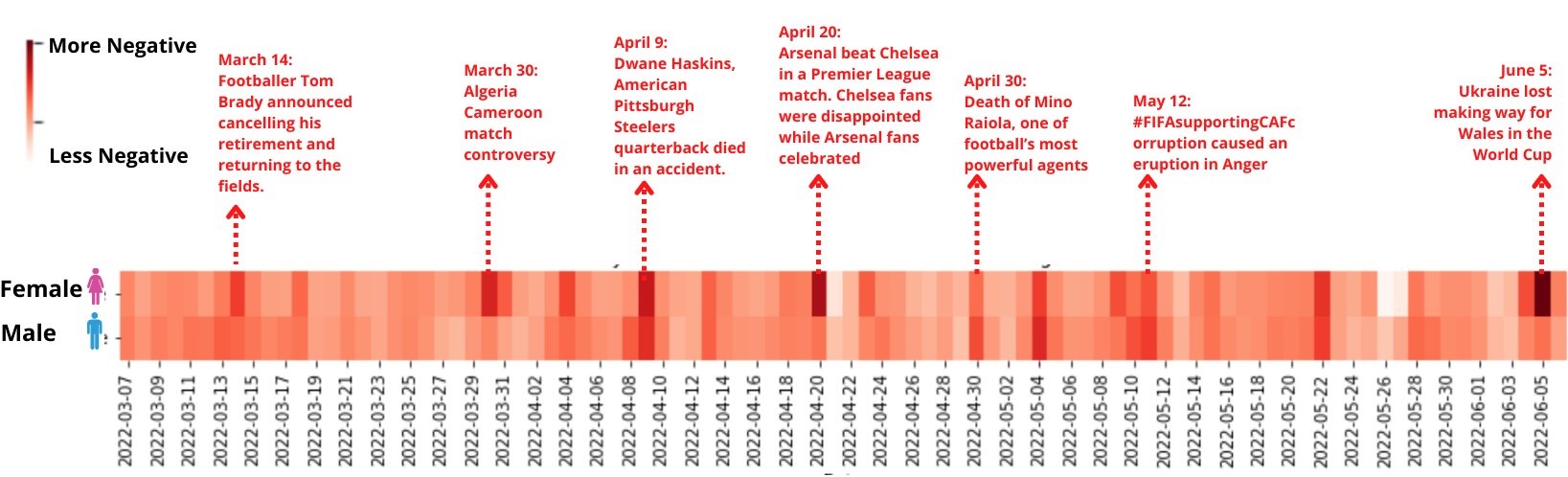}
         \caption{negative sentiment}
         \label{fig:five over x}
     \end{subfigure}
        \caption{Day wise distribution of VADER sentiment scores across male and female tweets (for English)}
        \label{appendixfigsentiment_distribution}
\end{figure*}

In Figure \ref{appendixfigsentiment_distribution}, we show sentiment analysis for English data categorized based on gender. We observed that the intensity of sentiments was different for both genders, and women displayed higher intensity. We also highlight events corresponding to peaks of positive and negative sentiments, and the same events can also be connected with emotion analysis. 

\section*{Appendix B}
\begin{figure*}[]
\centering
\includegraphics[width=\linewidth]{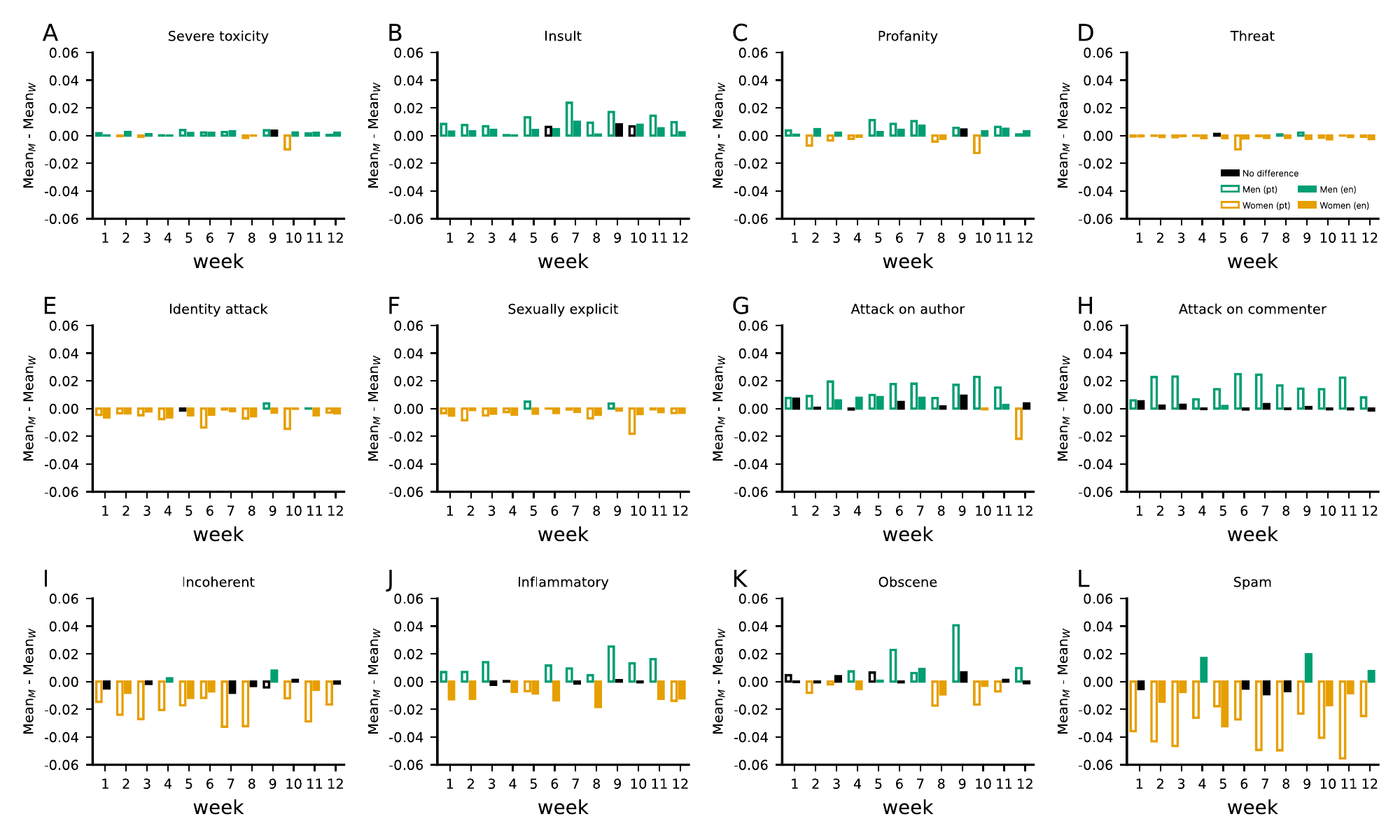}
\caption{Gender-based differences for different parameters computed using tweet text} 
\label{appendix_Figure_toxicity_2}
\end{figure*}

In Figure \ref{appendix_Figure_toxicity_2}, we show the plots for all parameters computed using Google Perspective API to understand the different dimensions of the content of the tweets. Toxicity is manifested in tweets that are impolite, disrespectful, or unreasonable, with the potential to drive individuals away from a discussion. Severe toxicity means a highly hateful, aggressive, or disrespectful comment, designed to strongly discourage a user from participating in a discussion or expressing their perspective. Profanity shows the use of swear words, curse words, or other obscene or profane language. As we discussed in the manuscript, men are more toxic and also have high severe toxicity. Besides toxicity, men write in such a way that it seems more attacking on the author and commenter of the tweet. Men also post more insulting content; an insulting, inflammatory, or negative comment directed towards an individual or a group of people is considered for computing the insult. Men use more abusive and swear words in both languages, displayed in profanity.

A threat involves expressing an intention to cause harm, injury, or violence against an individual or a group. There is no significant difference in using threatening language. Identity attack refers to making negative or hateful comments that specifically target someone based on their identity. Women are also very incoherent in the posted content as compared to males. The incoherent text means that it is difficult to understand and nonsensical. If the tweet text is irrelevant, then it is considered spam. Women's tweets are identified more as spam. Sexually Explicit refers to content that contains references to sexual acts, body parts, or other lewd material, and women raise such points more in online communication. The difference is not significant for inflammatory and obscene text. Inflammatory text intends to provoke or inflame. Obscene refers to the use of obscene or vulgar language, including cursing.

The way men and women tweet and present their opinion and emotions tend to be consistent in both the languages.

\end{document}